\documentclass[12pt]{iopart}

\usepackage{graphicx}%
\usepackage{multirow}%

\expandafter\let\csname equation*\endcsname\relax

\expandafter\let\csname endequation*\endcsname\relax

\usepackage{amsmath,amssymb,amsfonts}%
\usepackage{mathrsfs}%
\usepackage[title]{appendix}%
\usepackage{xcolor}%
\usepackage{textcomp}%
\usepackage{manyfoot}%
\usepackage{booktabs}%
\usepackage{algorithm}%
\usepackage{algorithmicx}%
\usepackage{algpseudocode}%
\usepackage{listings}%
\usepackage{hyperref}
\usepackage{pgf}
\usepackage[backend=bibtex, sorting=none]{biblatex}
\begin{document}

\title[Frustrated Ising Model on D-Wave]{Modeling a frustrated Ising square lattice \\with the D-Wave Quantum Annealer}

\author{C Marin$^{1,2}$, A Fontana$^3$, V Bellani$^{3,4}$, F Pederiva$^{5,6}$, A Quaranta$^{7,6}$, F Rossella$^{3,8}$, A Salamon$^{2}$, G Salina$^2$}

\address{$^1$ Department of Physics, University of Rome Tor Vergata, Rome,
00133, Italy.}
\address{$^2$ Tor Vergata section, INFN, Rome, 00133, Italy.}
\address{$^3$ Pavia section, INFN, Pavia, 27100, Italy.}
\address{$^4$ Department of Physics, University of Pavia, Pavia, 27100, Italy.}
\address{$^5$ Department of Physics, University of Trento, Trento, 38123, Italy}
\address{$^6$ INFN-TIFPA Trento Institute of Fundamental Physics and Applications, Trento, 38123, Italy}
\address{$^7$ Department of Industrial Engineering, University of Trento, Trento, 38123, Italy}
\address{$^8$ Department of Physics, Informatics and Matematics, University of Modena and Reggio Emilia, Modena, 41125, Italy.}

\vspace{10pt}

\begin{abstract}
The Ising model with nearest-neighbor interactions on a two-dimensional (2D) square lattice is one of the simplest models for studying ferro-magnetic to para-magnetic transitions. 
Extensive results are available in the literature for this model, which has become a paradigm for the study of magnetic phase transitions in materials, both theoretically and numerically. After a brief review of the main results obtained with a classical computer, we show how to implement on the D-Wave quantum annealer a more complex Ising model with the addition of competing antiferromagnetic interactions between the diagonal next-to-nearest neighbors with two coupling constants $J_1$ and $J_2$. The dynamics of this system, owing to frustration, are richer than those of the simple Ising model and exhibit a third striped (or antiferromagnetic) phase in addition to the ferro- and para-magnetic phases. This situation is similar to that observed in real 2D materials, in which competing interactions introduce new physical behaviors, such as magnetic or topological phase transitions in mono- or bilayer graphene. In this work, we observed all three phases on the D-Wave hardware, studied the behavior of the solution with different annealing parameters, such as the chain strength and annealing time, and showed how to identify the phase transition by varying the ratio between the ferromagnetic and antiferromagnetic couplings. The same system is studied on a classical computer, with the possibility of taking into account the temperature (fixed on D-Wave) as a free parameter and to explore the full phase diagram: some comparative conclusions with D-Wave are drawn. 
\end{abstract}

%
\vspace{2pc}

\section{Ising model and phase transitions}
The study of phase transitions plays a central role in modern physics in different contexts, from high-energy interactions at the LHC where the Quark-Gluon Plasma (QGP) is investigated to try to understand the very first moments after the Big Bang \cite{QGP2017} to innovative materials, where both magnetic and topological phase transitions are sought to open new possibilities for emerging applications, such as Quantum Technologies (QT) and new computing techniques. The topic is also becoming of interest in other disciplines, for example in medicine, in which the spread of diseases, such as cancer \cite{Davies2011} or Covid-19 \cite{Sole2021} is analyzed as a phase transition of macroscopic and complex systems, such as cities or countries. These are just a few emerging examples that show the great efforts currently devoted to the study of different systems with a unified view and method, which motivates the need for further studies in the field. See \cite{Baxter2007} for a detailed account of this subject.

\noindent The modeling of phase transitions is the first step in the study of any system, and a boost in the theoretical approach to these studies was certainly given by the introduction of the \textit{Ising model} \cite{1925Ising}. This model was introduced about a century ago and has become a paradigm for the study of phase transitions in magnetic systems, but not only. The seminal idea led in the following years, also thanks to the existence of the analytical solution of the problem in 1D and in parallel with the development of modern computers and algorithms that allowed the solution of the 2D problem to a variety of different models that are currently the well-defined research field of the so-called \textit{complex systems}.

\noindent We recall the main results for the most common version of the Ising problem found in the literature, that is, the square lattice in 2D with nearest-neighbor ferromagnetic interactions, which is described by the following Hamiltonian: 
\begin{equation}
H = -J \sum_{\langle i,j \rangle} s_i^z s_j^z + \mu_0 B \sum_{i} s_i^z 
\end{equation}
where $J$ is the coupling constant of the interaction, the first sum is extended to the nearest-neighbors only $\langle i,j \rangle$ and $B$ is an external magnetic field that in the following will not be considered (i.e. $B=0$).
This system represents the simplest model of a magnet that exhibits a phase transition from a ferromagnetic state to a paramagnetic state of a spin system at a critical temperature $T_c$ which is known exactly. This temperature is commonly reported as $k_B T_c/J=2/ln(\sqrt{2}+1)=2.26918...$ and can be easily calculated using a Monte Carlo simulation. The importance of this model is due to the fact that it is one of the few interacting models that have been solved analytically by Onsager \cite{Onsager1944}, who found the expression of its partition function. Therefore, the analytical and numerical solutions of the Ising model are important landmarks in the field of statistical mechanics and have significantly influenced our understanding of phase transitions in general.

\noindent
This result is well applicable to real materials, particularly to 2D materials, such as single- and bi-layer graphene and single-layer transition metal dichalcogenides (TMDs). Our method can be applied to understand spin-order transitions. For example, in bilayer graphene in the quantum Hall regime, a high-mobility sample exhibits an insulating state at the neutrality point, which evolves into a metallic phase when a strong in-plane field is applied, as expected for the transition from a canted antiferromagnetic to a ferromagnetic spin-ordered phase \cite{pezzini2014critical}. Another quantum phase transition that can be studied using methods that exploit our approach could be the plateau-insulator quantum phase transition, which has been experimentally explored in graphene \cite{amado2010plateau}. This transition is characteristic of several other 2D systems in the framework of the scaling theory of the integer quantum Hall effect (IQHE) and can be interpreted as a quantum phase transition with associated universal critical exponents \cite{amado2010plateau}. A great effort is currently underway to study magnetic or topological phase transitions in innovative materials using Quantum Computers, particularly Quantum Annealers. 
In this case, the theory is based on the Hubbard model {\cite{Gutzwiller1963, Kanamori1963, Hubbard1964}} and its extensions. In particular, the Kane-Mele-Hubbard model \cite{Hohenadler2012} which is able to describe topological transitions in graphene flakes \cite{Racz2020}, is based on a Hamiltonian inspired by the Ising model with competing interactions (Coulomb and spin-orbit). The dynamics of the system are analogous to those of the system studied here, but the mapping of such Hamiltonians into an Ising-like form suitable for implementation on a quantum annealer is a new, emerging, and non-trivial task. Some guidelines, based on the transformation from fermionic operators to spin operators (for example, the Jordan-Wigner or Bravyi-Kitaev transformations) and on the restriction to two-body interactions only, have been applied to simple molecules \cite{Xia2017, Levy2022}. However, to date, a general technique that could be applied to more complex systems has not yet been defined, although in \cite{Levy2022} the use of ancillary qubits in addition to those of the original lattice is applied to a Fermi-Hubbard problem on a 6x6 square lattice.

\subsection{Frustrated Ising model}

\begin{figure}[!htb]
\begin{center}
\includegraphics[scale=0.2]{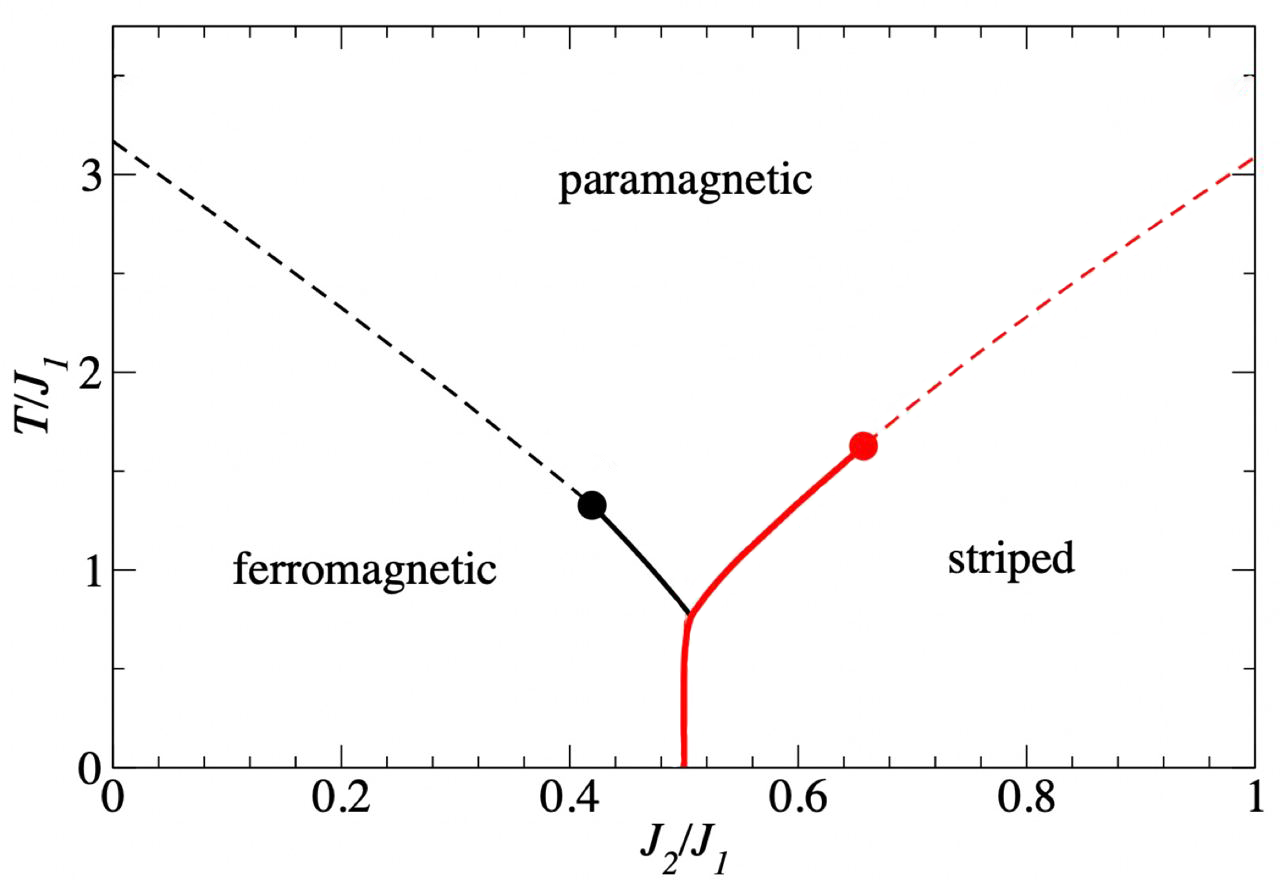}
\caption{Phase diagram calculated with the Mean Field approximation \cite{Jin2013}. It is possibile to see the different groud states changing the model's parameters. At low temperatures, the system has two different ground states depending on the ratio $J_2/J_1$: if it is below $0.5$, the ground state is ferromagnetic; otherwise it is striped. At high temperatures, the system shows paramagnetic states. }
\label{fig1}
\end{center}
\end{figure}

Adding competing interactions to the traditional Ising model provides a route
for the appearance of new phases and, in some cases, new types of phase
transitions. The simplest way to incorporate frustration in the standard Ising
model is to include next-nearest-neighbor antiferromagnetic interactions.
The Hamiltonian can be written as
\begin{equation}
H = -J_1 \sum_{\langle i,j \rangle} s_i^z s_j^z  + J_2 \sum_{\langle \langle i,j \rangle \rangle} s_i^z s_j^z + \mu_0 B \sum_{i} s_i^z 
\label{eq2}
\end{equation}
where $J_1$and $J_2$ are the two coupling constants acting on the nearest-neighbors ($\langle i,j \rangle$) and on the next-nearest-neighbors ($\langle\langle i,j \rangle\rangle$) along the diagonals. In addition,
in this case, the magnetic field was switched off in all calculations.
Owing to the presence of a frustrated interaction, this system presents
 richer dynamics with respect to the simple Ising model, depending not only on the temperature but also on the ratio $r=J_2/|J_1|$: in which a new
phase appears in which the spins align themselves with a striped pattern.
The system is well known and several studies on its phase transitions can be found in the literature \cite{Kalz2008, Jin2013, Sadrzadeh2018}. The system cannot be solved analytically, but different techniques, such as the Monte Carlo method or a Mean Field Theory (MFT) approximation, can be used to calculate an approximate phase diagram (see Fig. \ref{fig1} which refers to an MFT calculation).

\noindent In this paper, we study the phase transitions of this system with a fixed lattice size (20x20) on a D-Wave Quantum Annealer. This approach is particularly interesting because this device natively solves problems in the Ising form. In addition, we study the limitation of this new technology by tracking the resources needed to solve different lattice sizes of the frustrated Ising model, and we will show how to physically transition the D-Wave hardware into the striped phase. Similar results have also been found in a recent study \cite{Park2022}. The goal of this work is to explore the potential of this new computing approach on a known system and to understand its limitations in view of its possible applications to more complex optimization problems.

\section{Quantum Annealer}
A quantum annealer is a lattice composed of qubits identified by Josephson junctions\cite{ManifacturingSpins}, which can interact with other qubits through programmable couplers. The D-Wave Quantum Annealers have different architectures, each of which differs by the number of couplers associated with each qubit. In this work, we used the Pegasus and Zephyr architectures\cite{PegasusTopology, ZephyrTopology}, which have different properties, as shown in Table \ref{tab:QPU_Architectures}.

\begin{table}[h]
    \centering
    \begin{tabular}{ccc}
        Architecture & Total Qubits & Connectivity Degree\\
        \hline
        Pegasus & 5627 & 15\\
        Zephyr & 563 & 21\\
    \end{tabular}
    \caption{Summary of the characteristics for different D-Wave architectures. The column of total qubits represents the number of total qubits in a QPU with a determined architecture, and the connectivity degree represents the number of couplers associated with each qubit. }
    \label{tab:QPU_Architectures}
\end{table}
\noindent
\newline  The state of a qubit can be in a superstate position of states 0 and 1, and under normal conditions the probability of falling in one of the classical states is equal (50\%). The quantum annealer can be described by the following Hamiltonian:
\begin{equation}\label{eq:DWaveHamiltonian}
    H = \underbrace{-\frac{A(t/t_{max})}{2} \sum_i \hat{\sigma}_x^{(i)}}_\text{Initial Hamiltonian} + \underbrace{\frac{B(t/t_{max})}{2}\left ( \sum_i h_i \hat{\sigma}_z^{(i)} + \sum_{i>j} J_{ij} \hat{\sigma}_z^{(i)}\hat{\sigma}_z^{(j)} \right )}_\text{Final Hamiltonian}
\end{equation}
where $\hat{\sigma}_{x, z}^{(i)}$ are Pauli matrices operating on qubit $\mathbf{q_i}$ and $h_i$ and $J_{ij}$ are the qubit biases and coupling strength, respectively. The Hamiltonian is the sum of two terms: the \textit{initial Hamiltonian} and \textit{final Hamiltonian}. The first represents the lowest-energy state of the system when all the qubits are in a superposition state of 0 and 1, and the second is the lowest-energy state, which represents the result of the encoded problem. During a quantum annealing process \cite{QCAE, Childs_2001}, the system begins in the lowest energy eigenstate of the initial Hamiltonian, and as it anneals, it introduces the final Hamiltonian, which contains the biases and couplers, and reduces the influence of the initial Hamiltonian. At the end of the annealing, it is in an eigenstate of the Hamiltonian problem, and each qubit is in a classical state of 0 or 1. A Quantum Processor Unit (QPU) can solve binary optimization problems: the Ising model\cite{Lucas_2014} and Quadratic Unconstrained Binary Optimization\cite{glover2019tutorial} (QUBO). The Ising formulation is expressed by the discrete variables $s_i$, which correspond to the values of $\pm1$, and the objective function to be minimized is
\begin{equation}\label{eq:IsingFunc}
    E_{Ising}(\mathbf{s}) = \sum_{i=1}^N h_is_i +\sum_{i=1}^N\sum_{j=i+1}^NJ_{i,j}s_is_j
\end{equation}
\begin{figure}[!htb]
\begin{center}
\includegraphics{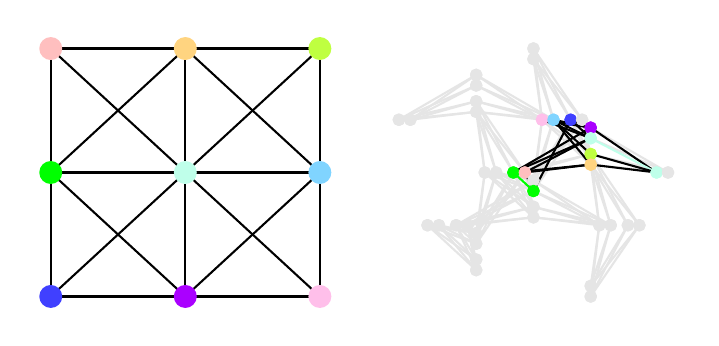}
\caption{Example of the embedding of a 3x3 lattice on the D-Wave Pegasus architecture. It is possible to see that some original qubits (the green and the light blue) are represented on the QPU using respectively 2 qubits each through the colored chain links.}
\label{fig2}
\end{center}
\end{figure}
where the linear coefficient $h_i$ corresponds to the qubit bias and the quadratic coefficient $J_{i, j}$ corresponds to the coupling strength (its sign can correlate the states of the variables or anti correlate them).
On the other hand, QUBO problems use Boolean variables $x_i$, taking values of 1 (true) and 0 (false). These formulations can be translated into the form of a graph that comprises all the information in a collection of nodes and edges. The problems must be embedded in the QPU architecture by mapping the logic variable with a qubit or a chain of qubits if the links of a qubit cannot reproduce the interactions of the problem (these qubits must act as a single variable). Fig. \ref{fig2} shows to the left the lattice of the frustrated Ising model and to the right its embedding in the QPU architecture, where it can be seen that the embedded graph has two logical variables mapped into the QPU using two qubits each. \begin{figure}[!htb]
\begin{center}
\includegraphics[scale=0.8]{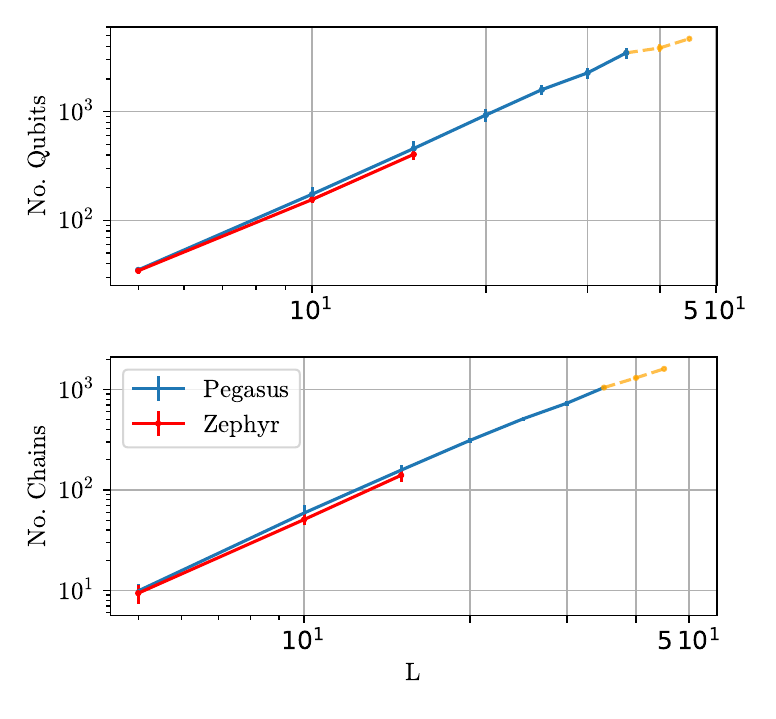}
\caption{Qubits and Chains used for embedding at various lattice size. As discussed in the text, both D-Wave architectures provide solutions only up to a maximum size of the frustrate ising lattice, $L=15$ for the Zephyr architecture and $L=35$ for the Pegasus architecture. On Pegasus, the embedding process starts to fail when the lattice size is more than 40, at 40 we obtain a success ratio 95\%, while at 45 we obtain a solution 10\% of the times. Further incresing the lattice size, no solutions are found.}
\label{fig11}
\end{center}
\end{figure}
This process was extended to different lattice sizes and QPU topologies, to investigate the resources required to solve the problems. In Fig. \ref{fig11}, we see an exponential growth of the number of qubits and the number of chains needed to represent the frustrated Ising model in the QPU. This shows one of the limitations of the quantum annealer, because the QPU can be easily saturated by relatively small size problems (\cite{camino2023quantum, Xia2017}). Furthermore, different topologies require different resources to map a problem on the QPU, and this is explained by the connection degree of each qubit: if a qubit has a sufficiently high amount of links (as in the Zephyr topology), it is easier to find an embedding by using less resources by the fact that Zephyr curves are always below the Pegasus curve (Fig.\ref{fig11}).

\subsection{Annealing parameters}
The problems submitted to D-Wave quantum solvers include problem parameters (e.g., the linear coefficient h or the quadratic coefficients of an Ising problem) and solver parameters that control how the problem is executed (e.g., annealing time and anneal\_schedule). Some of these parameters are fundamental for solutions because they ease the exploration and exploitation of the solution space, particularly the chain\_strength and  annealing\_time parameters. 

\subsection{Chain strength}
\begin{figure}[!htb]
\begin{center}
\includegraphics[scale=0.80]{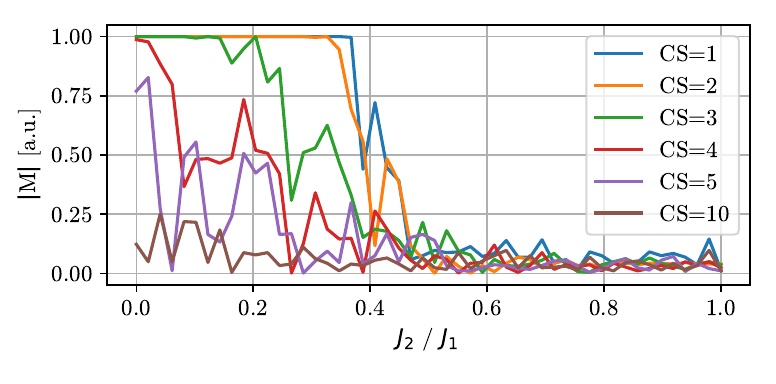}
\caption{Mean absolute magnetization of 20x20 system for different values of the chain strength. It is shown that the QPU cannot find the correct results when this values is grater than 1.}
\label{fig3}
\end{center}
\end{figure}
\noindent The chain technique was introduced to represent a problem variable as multiple qubits were chained together. To do so, all of these qubits must return the same value for a sample, which is accomplished by setting a strong coupling to the edges connecting these qubits. The problem constraints are rescaled on the QPU in the range of -1 to 1 for both biases and coupler strengths. Taking into account the ratio $J_2/J_1$ of the frustrated Ising model and iterating on different values of chain\_strength, it is possible to plot Fig. \ref{fig3}, which shows how the values of chain strength greater than the maximum weight of the problem have a negative influence on the results. The frustrated Ising model has a phase transition for J2/J1 = 0.5, and the QPU reproduces this result for a chain strength value equal to 1. When these values increase, the QPU can no longer reproduce the theoretical results. What happens when chain strengths are too small or too large? If the parameter is too small, the physical qubits in the chain do not assume the same value at the end of the annealing process, and the chain breaks. If the chain breaks, the solutions returned from QPU may be degraded and suboptimal. If the chain strengths are greater than the largest absolute value in QUBO, the terms related to the chain strengths are set to 1, and the QUBO weights are proportionally smaller. As the chain strength increases, the individual QUBO terms and problem constraints shrink to near zero. This allows each chain to act as a separate entity, and the initial QUBO problem transforms into a problem with N independent variables that do not interact.

\subsection{Annealing time}
\begin{figure}[b]
\begin{center}
\includegraphics[scale=0.8]{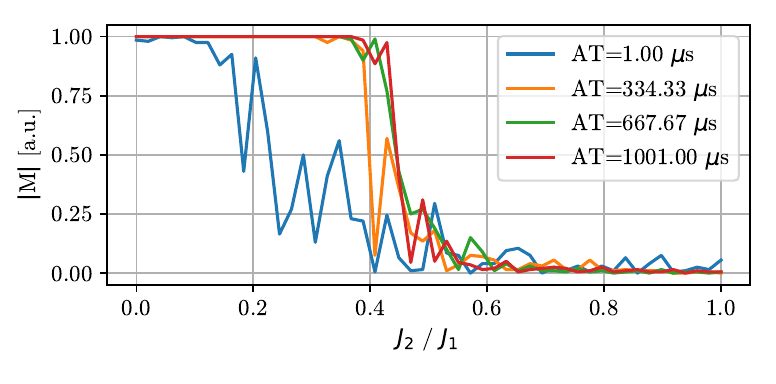}
\caption{Effect of the annealing time on the transition plot of the Frustrated Ising model. It is shown that for small annealing times, the result is noisy and differs from the results given in the literature, while for large times the line is smoother, and the transition point is close to the theoretical one. }
\label{fig4}
\end{center}
\end{figure}
\noindent Annealing time is a parameter of the quantum processor that defines the duration of a run. Its default value is equal to 20 $\mu s$ and the quantum processor used in this work ( Advantage 4.1 QPU) has a resolution of 0.01$\mu s$. Taking into account the mean magnetization and changing the value of the ratio $J_2 / J_1$, in Fig.
\ref{fig4}, it is possible to see the effect of the annealing time on the solution found by the QPU. For small annealing times, the transition point is translated to the left with respect to the theoretical value, and the plot is noisy. When the annealing time increased, the magnetization became smoother, and the transition point approached the theoretical value. The optimal value of this parameter is a trade-off between the annealing time and the number of reads requested, because the time limit of the entire system is set to 1 $ms$. If a simulation exceeds this limitation, the QPU preprocessor raises an error.

\section{Phase diagram analysis}

Phase transitions in physics are a very interesting and challenging topic.
The phase of a system is defined by an order parameter, that quantifies 
the degree of organization of the system, and a transition typically 
occurs when a system undergoes a change of state. Typical examples are 
the state change of a material from solid to liquid to gas by changes 
in pressure or density or the magnetization of a material in which the 
phase can be characterized by the net alignment of spins.
A phase transition is characterized by a sharp transition in the value of
the order parameter, which occurs at the critical value of a control
parameter that influences the state of the system. Classical
examples of control parameters include the temperature or strength of an 
applied external magnetic field. In general, the order parameter operates 
as a dependent variable, while the control parameter acts as an independent 
variable, and a plot of the order parameter as a function of the control parameter 
 exhibits a discontinuity or singularity at the critical value of the control
parameter, which is the key to understanding the underlying dynamics
of the transition process. In our case, the control parameters are the temperature T and the ratio $r=J_2/|J_1|$, whereas for the order parameter, we analyzed two different observables in the lattice: the magnetization and the domain wall length.

\begin{figure}[!htb]
\begin{center}
\includegraphics[scale=0.6]{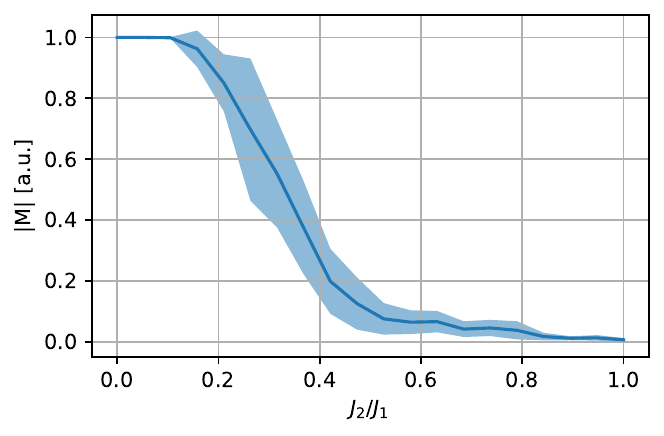}
\includegraphics[scale=0.0675]{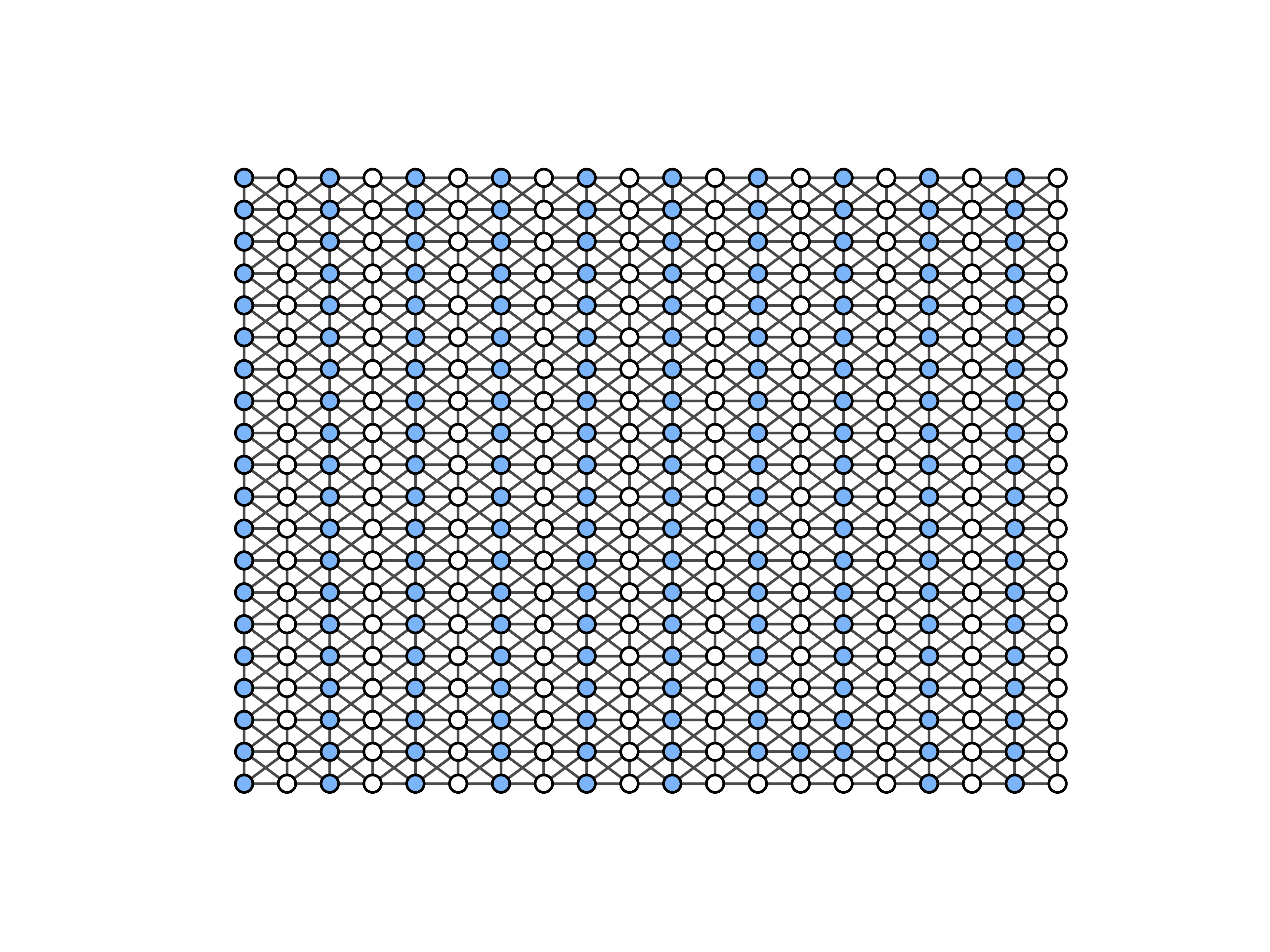}
\caption{ On the left is shown themagnetization plot for the L=20 lattice, the blue line is the mean value of the magnetization, and the blue area the represents its standard deviation. On the right, it is shown the striped configuration obtained on the D-Wave lattice (right).}
\label{fig5}
\end{center}
\end{figure}

\noindent In the first study, we evaluated the magnetization of a 20x20 lattice to investigate the transition from the ferromagnetic phase to the striped phase, as shown in Fig. \ref{fig1}. By performing a sweep on the r parameter, starting from a ferromagnetic configuration ($r=0$), we obtained the magnetization plot shown in Fig. \ref{fig5} (left). Inspection of the D-Wave qubit lattice for $r=1$ reveals the transition of the system to the striped phase, as shown in Fig. \ref{fig5} (right).

\noindent The inflection point in the transition curve occurs at $r\approx 0.3$, indicating that the temperature of the D-Wave lattice is above the triple-point temperature in the theoretical phase diagram. In this case, the transition also passes through an intermediate paramagnetic phase with a checkerboard-like lattice pattern. To better understand and discuss this result, we performed the same calculations on a classical computer with a Monte Carlo program based on the Metropolis algorithm \cite{Metropolis_1953}. This extra calculation is relevant in the study of the D-Wave lattice magnetization, because on this device, the temperature T cannot be directly controlled; only the classical calculations can provide access to 
the full phase diagram in the (r,T) plane, and we will use the results obtained in this way to directly compare the behavior of the different order parameters to estimate the lattice temperature on the D-wave.

\subsection{Magnetization phase diagram}

\begin{figure}[!htb]
\begin{center}
\includegraphics[scale=1]{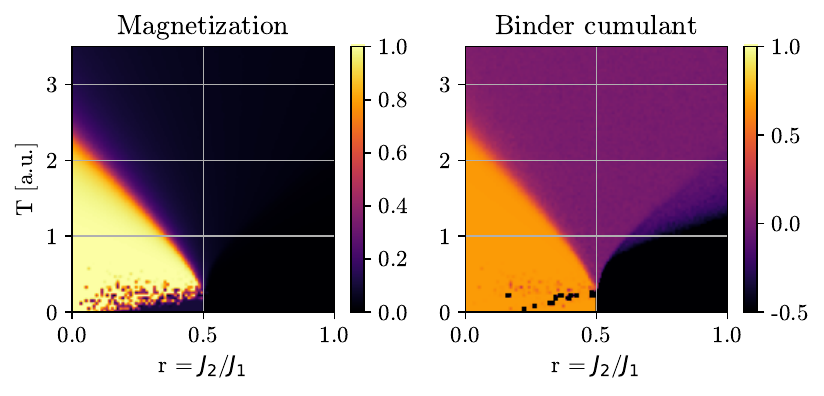}
\caption{Phase Diagrams of the Ising model with competing interactions: Magnetization left, Binder cumulant right. The phase diagram of both quantities shows some noise at low temperatures and near the transition point $r=J_2/J_1$, this is due to the computation of the exponential during the Monte Carlo simulations. The phase diagram of the magnetization does not show the three phases, while the Binder's cumulant shows clearly the three ground states of the system.}
\label{fig6}
\end{center}
\end{figure}

The overall magnetization of the lattice can be calculated as the average of the lattice spins for each configuration in the (r,T) plane:
\begin{equation}\label{eq:magnetization}
    M = \frac{1}{L^2} \sum_i s_i
\end{equation}
The result is shown in Fig. \ref{fig6} (left) for a Monte Carlo calculation with 2000 histories. The ferro- and paramagnetic phases, as expected from the MFT approximation calculations \cite{Jin2013}, are clearly visible, while the striped phase barely emerges in the high r-low T region of the diagram. 
This is basically due to the fact that both the para-magnetic and striped phases correspond to a net zero magnetization of the lattice, and this observable is not able to distinguish between the two spin configurations. Therefore, we also investigated other quantities to improve the analysis.

\subsection{Binder cumulant phase diagram}

Binder's cumulant is an interesting statistical variable that can better distinguish between paramagnetic and striped phases. It is a quantity introduced in the context of finite-size scaling that allows the critical point and exponents to be located \cite{Selke_2005}.  An Ising model with zero field is defined as:
\begin{equation}\label{eq:binder}
    U = 1 - \frac{1}{3} \frac{\langle M^4 \rangle}{\langle M^2 \rangle^2} 
\end{equation}
where $\langle M^4 \rangle$ and $\langle M^2 \rangle$ are the central moments of the fourth and second order of magnetization, respectively. In the thermodynamic limit, where the system size $L \rightarrow \infty$, $U \rightarrow 0$ for $T>T_c$ (because in the large L limit, $\langle M^4 \rangle=\langle M^2 \rangle ^2$) and $U \rightarrow 2/3$ for $T<T_c$ (as in this case, the order parameter is Gaussian, and all cumulants vanish except for the variance, leading to $\langle M^4 \rangle=3\langle M^2 \rangle$). Thus, the function is discontinuous in this limit, but its properties are also relevant for finite-size lattices; we exploit this property in our analysis. This quantity is much more effective than magnetization in identifying the striped phase, as can be seen in the phase diagram shown in Fig. \ref{fig6} (right). The three phases are well distinguishable, and the triple point of the system can also be clearly identified.

\subsection{Domains wall length phase diagram}

\begin{figure}[!htb]
\begin{center}
\includegraphics[scale=0.6]{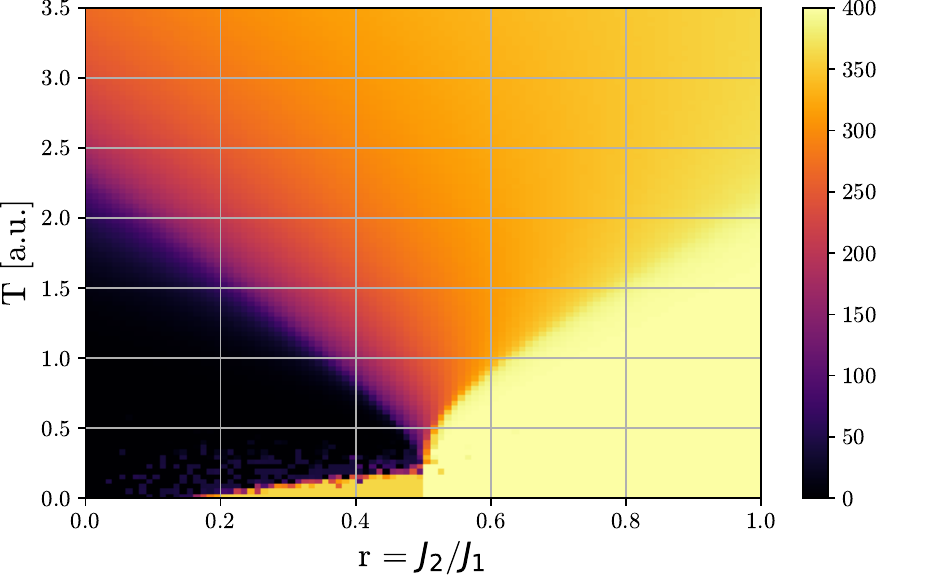}
\caption{Phase diagram using the wall domain lenght for the L=20 lattice. This quantity shows clearly the three ground states and the noise of the computation method is reduced at low temperatures. }
\label{fig8}
\end{center}
\end{figure}

In addition to the binder cumulant, we calculated the perimeter of the domain walls in the simulations using both classical and quantum computers. 
A domain wall is the boundary between two neighboring magnetic domains with opposite magnetizations present at any given step in the time evolution of the lattice. In our approach, the wall length is simply given by the total perimeter of the interface separating the two magnetic phases, and is intended with zero thickness. This new quantity is obtained by analyzing the lattice results at each step and counting the sites that belong to the boundary of the domains. 
The expression used in this work can be written as: 
\begin{equation}\label{eq:wall}
    W = \sum_{s_1=+1} l_i
\end{equation}
where $l_i$ is the perimeter of each domain and the sum is extended to all domains. A phase diagram obtained using this quantity is shown in Fig. \ref{fig8}.

\section{Discussion}

In this study, we found different results: 
\begin{itemize}
    \item We exploited the D-Wave embedding process at different lattice sizes to determine its actual limitations.
    \item We found the optimal D-Wave parameters to optimize the annealing process, to reproduce the classical behavior at the phase transition.
    \item We developed a classical Monte Carlo code to calculate the full phase diagram.
    \item We observed all three phases characterizing the frustrated Ising model with the annealer and the classical code.
\end{itemize}
From these results, we can now attempt to evaluate the lattice temperature by fitting the phase diagram profile as a function of the ratio r for any fixed temperature T: we obtain a series of profiles that can be compared with the profile obtained with the QPU of the D-Wave by means of the R-squared coefficient \cite{chicco_2021} to select the best match. 
The temperatures obtained are listed in Table \ref{tab:R-square}.

\begin{figure}[!htb]
\begin{center}
\includegraphics[scale=0.8]{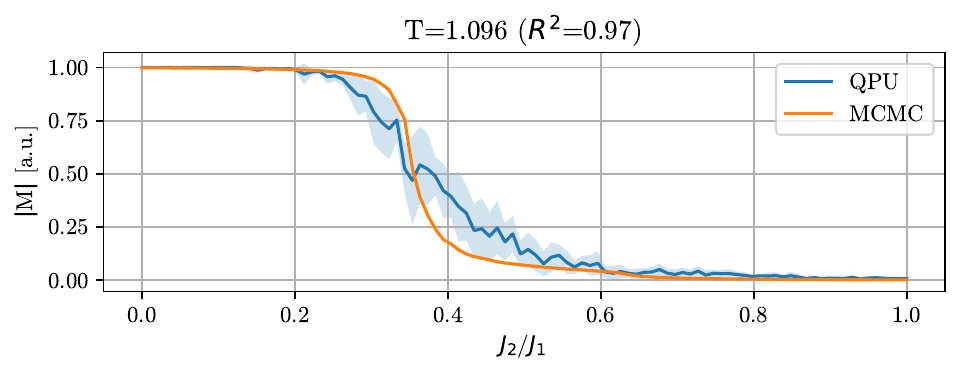}
\caption{Best fit of magnetization at fixed temperature.}
\label{fig7}
\end{center}
\end{figure}

\begin{figure}[!htb]
\begin{center}
\includegraphics[scale=0.8]{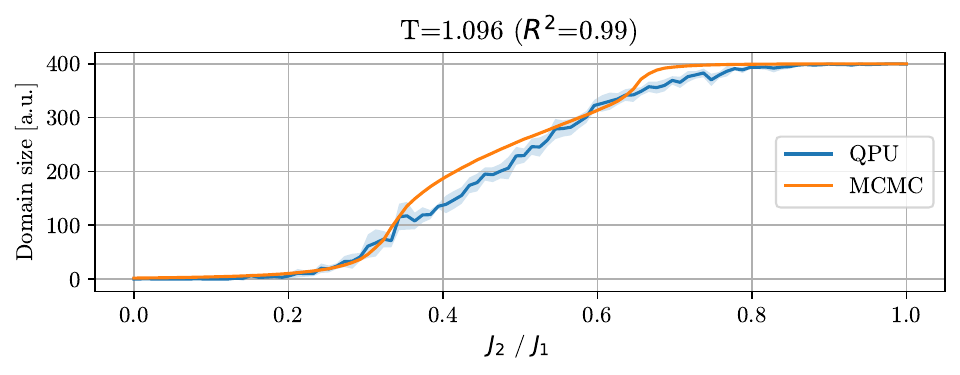}
\caption{Best fit of domain wall length at fixed temperature. }
\label{fig9}
\end{center}
\end{figure}

\begin{table}[]
\centering
\begin{tabular}{ccc}\hline
Quantity & T & $R^2$ \\  \hline
$M$ & 1.096 & 0.97\\  
$W$ & 1.096 & 0.99\\   \hline
\end{tabular}
\caption{Best-fit results for the estimate temperature of the lattice with the different observables considered.}
\label{tab:R-square}
\end{table}
\noindent The corresponding temperature provided an estimate of the lattice temperature in the D-Wave architecture. The profiles that exhibit the highest R-squared values obtained with magnetization and the domain wall are shown in Fig. \ref{fig7} and Fig. \ref{fig9}. The fit was not performed on the phase diagram obtained from the Binder cumulant because of conceptual problems in its calculation on QPUs. In particular, it is not possible to calculate the fourth-order and second-order moments on a QPU because of the lack of a historical magnetization series during the annealing process.
From Fig. \ref{fig9}, it is also possible to notice that the R-squared values 
are higher than the magnetization value, and that the two lines were similar, with 
small differences near the transition point at r = 0.5.
We note that the temperatures obtained using magnetization (M) and the wall domain (W)
are the same, and considering the phase diagram in Fig. \ref{fig8}, it can be observed that the QPU profile passes through all phases.

\begin{figure}[!htb]
\begin{center}
\includegraphics[scale=0.9]{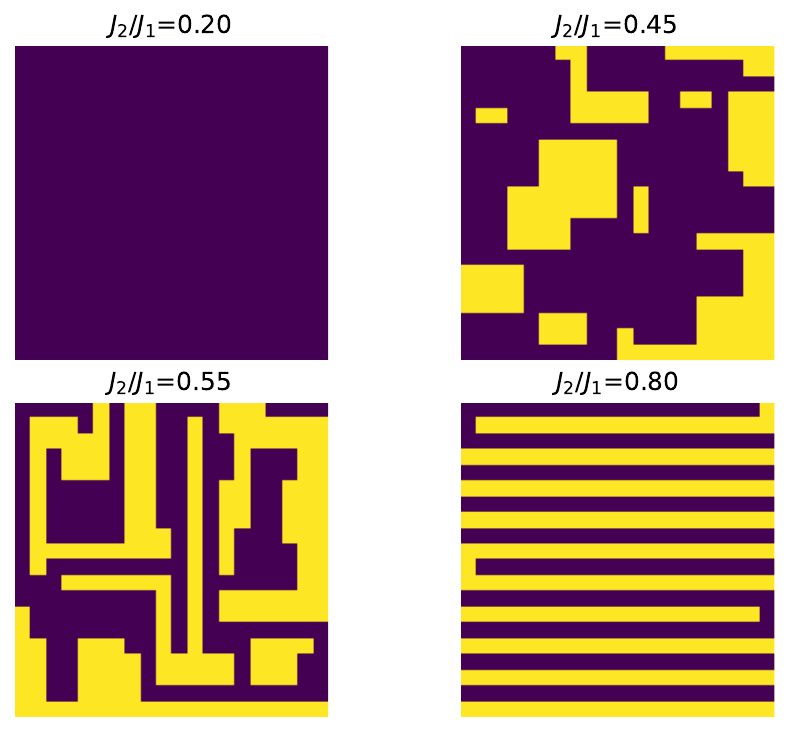}
\caption{Lattice magnetization patterns corresponding to the configurations 
at the lowest energy given by the QPU for different $J_2/J_1$ ratios.}
\label{fig10}
\end{center}
\end{figure}

\noindent Furthermore, from the phase diagram, it is possible to extract some relevant
values of $r=J_2/J_1$, specifically for the ferromagnetic and striped phases,
the chosen values are 0.20 and 0.80, while for the paramagnetic phase, the 
values 0.45 and 0.55 are chosen. 
In Fig. \ref{fig10}, the configurations with
the lowest energy for all these values are shown. It can be observed that
the QPU can reproduce all ground states of the Ising model with concurrent
interactions on the diagonal, in particular, the ferromagnetic phase and the
striped phase (with some artifacts due to thermal fluctuations). 
For values near the transition point $J_2/J_1$ = 0.5, it is possible to
notice that the configurations exhibit local behaviors of all three phases
because there are large domains that are characteristic of the ferromagnetic phase
and small domains composed of a few sites that characterize the paramagnetic
phase, which are arranged in stripes (the effect is more evident
at $J_2/J_1$ = 0.55).

\section{Conclusions}
In this work we studied a frustrated Ising model with next-to-nearest neighbors competing interactions on the Quantum Annealer D-Wave. 
This computer is based on a QPU architecture that uses a lattice of manufactured spin circuits realized with Josephson junctions and can solve problems presented in an Ising-like form. Therefore, the Ising model is an ideal system to begin working with this technology. 

\noindent The primary focus of this study is the characterization of quantum annealers using the frustrated Ising problem as a metric. The first results obtained relate to the embedding process of the model using different lattice dimensions, from which we found an exponential growth in the resources required to solve this problem on QPUs. This trend was found in both architectures (Pegasus and Zephyr), which exhibit a different degree of connectivity: the higher the degree of connectivity, the easier it is to find the embedding of the problem.

\noindent We also performed a tuning phase to define the best QPU parameters (chain strength and annealing time) to solve our problem. In this step, we also compared the results obtained for the same system on a classical computer with the Metropolis algorithm, to better understand the annealing process and assess its potential and limitations. In this optimized system, we later studied the phase transitions of the frustrated Ising model by analyzing two different observables: magnetization and the size of the magnetic wall domain. We estimated the lattice temperature on the QPU, and obtained consistent results with both methods. 

\noindent We conclude that the model simulated on the QPU passes near the triple point 
of the phase diagram, where the system exhibits three coexisting phases observed in the hardware. The Quantum Annealer is ideal for solving Ising-like problems and will also be a powerful tool for addressing more complex cases in physics, as the number of qubits and associated couplers will increase.

\ack 
We would like to thank the joint Quantum laboratory at Trento (Q@TN) and CINECA for granting us access to the D-Wave quantum annealer through the two-year activities ("Modeling many-electron states in nanowire quantum dots and 2D materials using the D-wave quantum annealer" and "Use of the D-Wave quantum annealer for optimization problems in complex systems").
All authors acknowledge the support of the INFN National Science Committee 5 for funding this activity.
A. Fontana, V. Bellani, A. Salamon acknowledge the support of PNRR MUR project PE0000023-NQSTI (Italy).

\section*{Compliance with ethical standards}
The authors declare that they have no conflict of interest.



\printbibliography

\end{document}